\documentclass[%
 reprint,
 amsmath,amssymb,
 aps,
]{revtex4-2}

\usepackage{graphicx}
\usepackage{dcolumn}
\usepackage{bm}

\begin{document}

\preprint{APS/123-QED}

\title{Growth and characterization of single crystal cubic TaN and hexagonal Ta$_2$N films on c-plane Sapphire}

\author{Anand Ithepalli}
\author{Amit Rohan Rajapurohita}
\author{Arjan Singh}
\author{Rishabh Singh}
\author{John~Wright}
\author{Farhan Rana}
\author{Valla Fatemi}
\author{Huili (Grace) Xing}
\author{Debdeep Jena}
\affiliation{Cornell University}

\date{\today}

\begin{abstract}

Two single crystal phases of tantalum nitride were stabilized on c-plane sapphire using molecular beam epitaxy. The phases were identified to be $\delta$-TaN with a rocksalt cubic structure and $\gamma$-Ta$_2$N with a hexagonal structure. Atomic force microscopy scans revealed smooth surfaces for both the films with root mean square roughnesses less than 0.3 nm. Phase-purity of these films was determined by x-ray diffraction. Raman spectrum of the phase-pure $\delta$-TaN and $\gamma$-Ta$_2$N obtained will serve as a future reference to determine phase-purity of tantalum nitride films. Further, the room-temperature and low-temperature electronic transport measurements indicated that both of these phases are metallic at room temperature with resistivities of 586.2 $\mu\Omega$-cm for the 30 nm $\delta$-TaN film and 75.5 $\mu\Omega$-cm for the 38 nm $\gamma$-Ta$_2$N film and become superconducting below 3.6 K and 0.48 K respectively. The superconducting transition temperature reduces with applied magnetic field as expected. Ginzburg-Landau fitting revealed a 0 K critical magnetic field and coherence length of 18 T and 4.2 nm for the 30 nm $\delta$-TaN film and 96 mT and 59 nm for the 38 nm $\gamma$-Ta$_2$N film. These tantalum nitride films are of high interest for superconducting resonators and qubits.

\end{abstract}

\maketitle




\begin{figure*}
    \centering
    \includegraphics[width=0.8\textwidth]{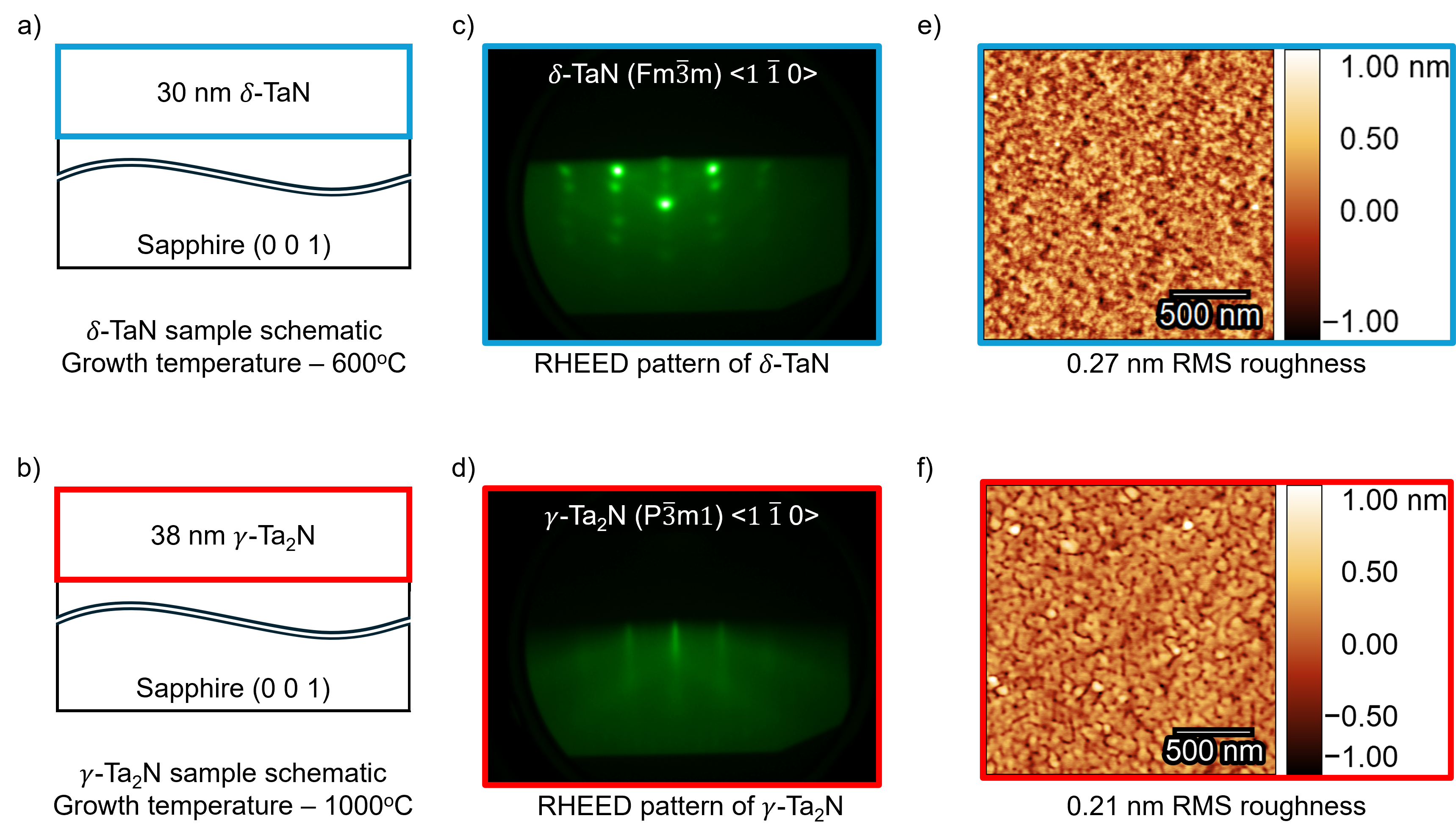}
    \caption{a) and b) Sample schematics of 30 nm $\delta$-TaN and 38 nm $\gamma$-Ta$_2$N films respectively. c) and d) RHEED patterns after the growth of the films shown in a) and b) with the space groups of the films and zone axes indexed according to the film orientations. These RHEED patterns indicate single crystal twinned, 3-dimensional growth mode of $\delta$-TaN and single crystal, 2-dimensional growth mode of $\gamma$-Ta$_2$N. e) and f) Atomic force microscopy (AFM) scans of the films shown in a) and b) indicate that both the films have smooth surfaces with root mean square (RMS) roughnesses below 0.3 nm.}
    \label{fig:RHEED_AFM}
\end{figure*}

\begin{figure}
    \includegraphics[width=0.5\textwidth]{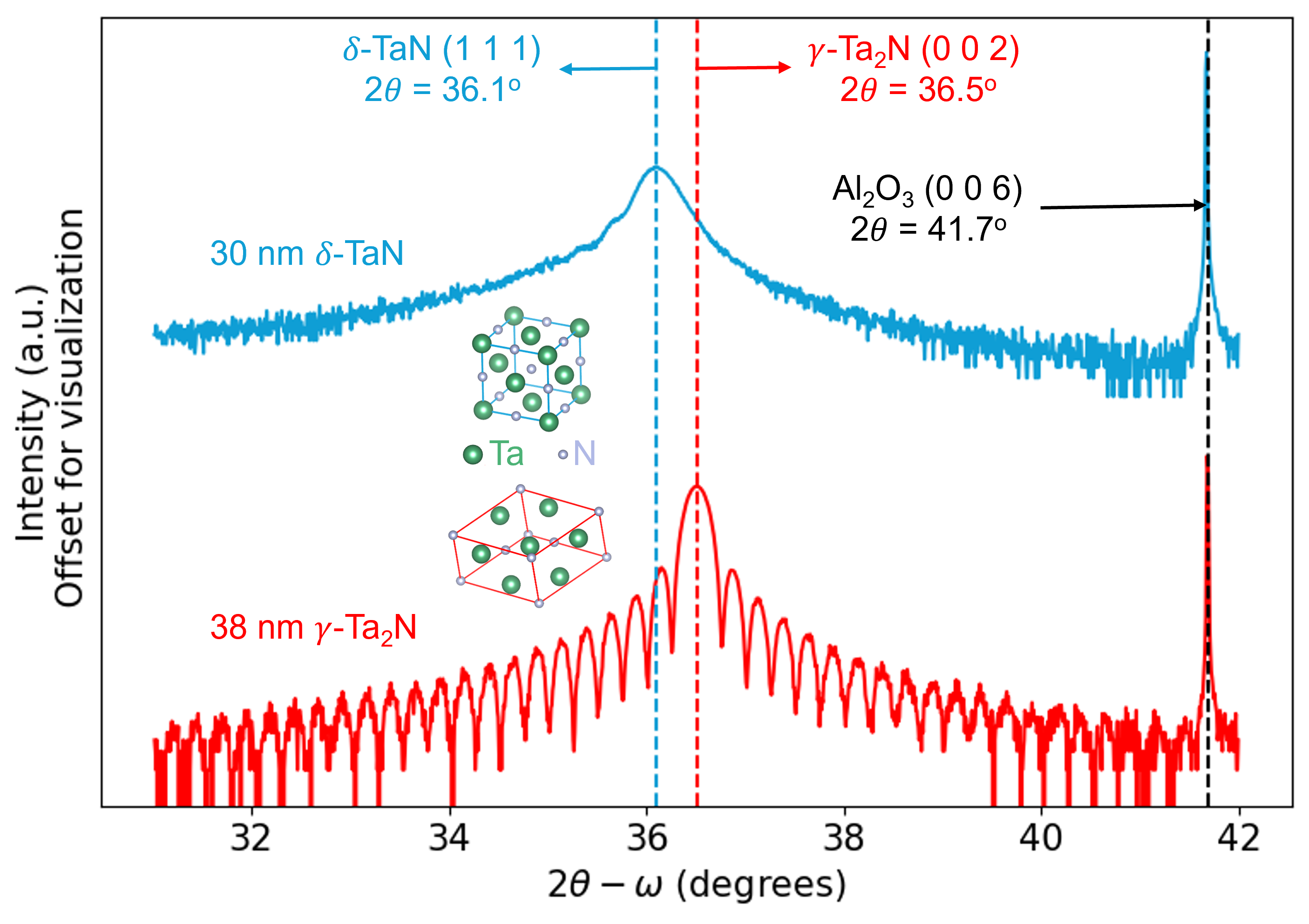}
    \caption{Singular peaks of the symmetric $2\theta$-$\omega$ coupled scans of the 30 nm $\delta$-TaN and 38 nm $\gamma$-Ta$_2$N films indicating single crystal nature of these films.}
    \label{fig:XRD}
\end{figure}

\begin{figure*}
    \centering
    \includegraphics[width=0.9\textwidth]{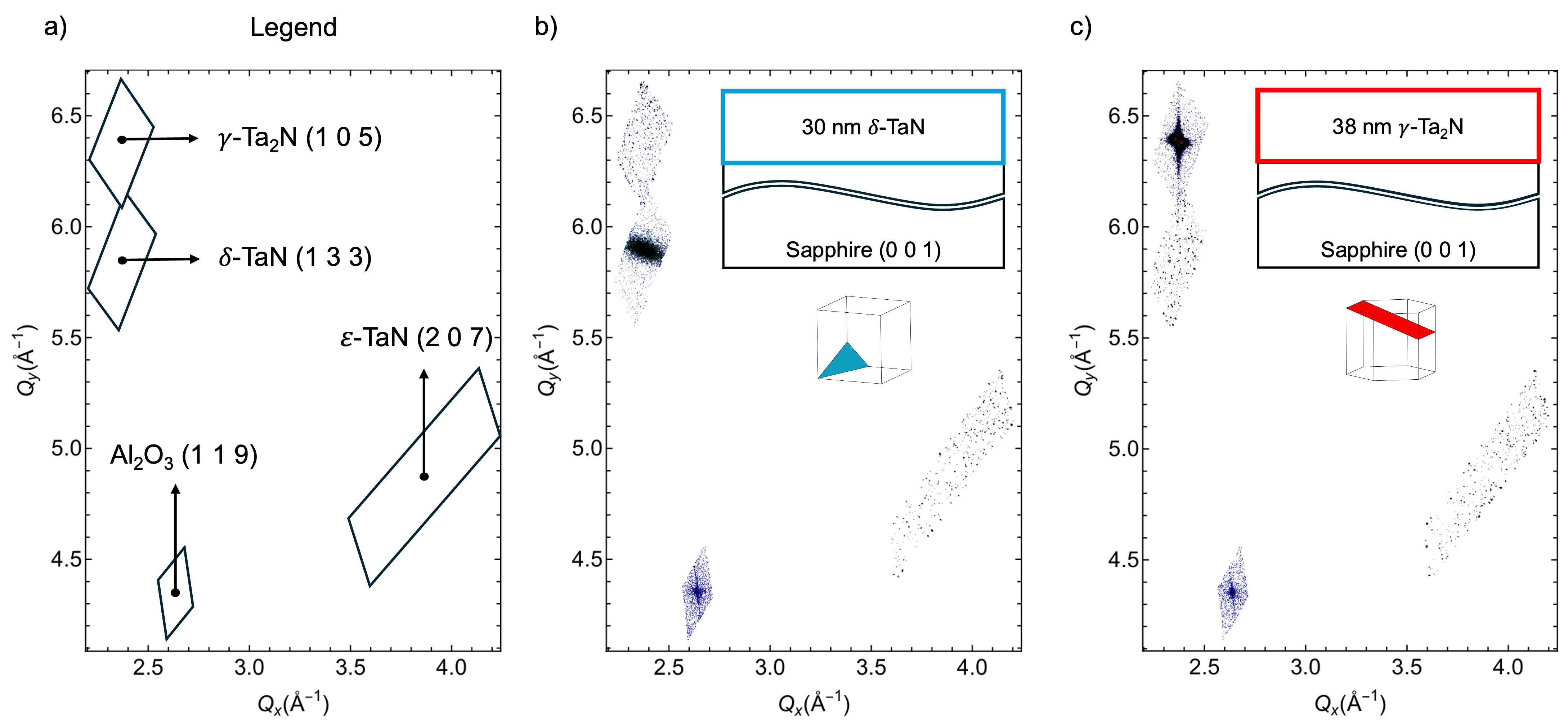}
    \caption{Reciprocal space mapping (RSM) figures are used to verify the presence of multiple phases and orientations of the films with respect to the substrate. a) Legend of the expected peak locations of the tantalum nitride phases of interest along side a sapphire substrate asymmetric peak. All the RSM figures shown here have planes corresponding to $<$\text{1 0 9}$>$ zone axis of the sapphire substrate. b) RSM figure of $\delta$-TaN and c) RSM figure of $\gamma$-Ta$_2$N indicating the presence of only intended phases.}
    \label{fig:RSM}
\end{figure*}

\begin{figure}
    \includegraphics[width=0.5\textwidth]{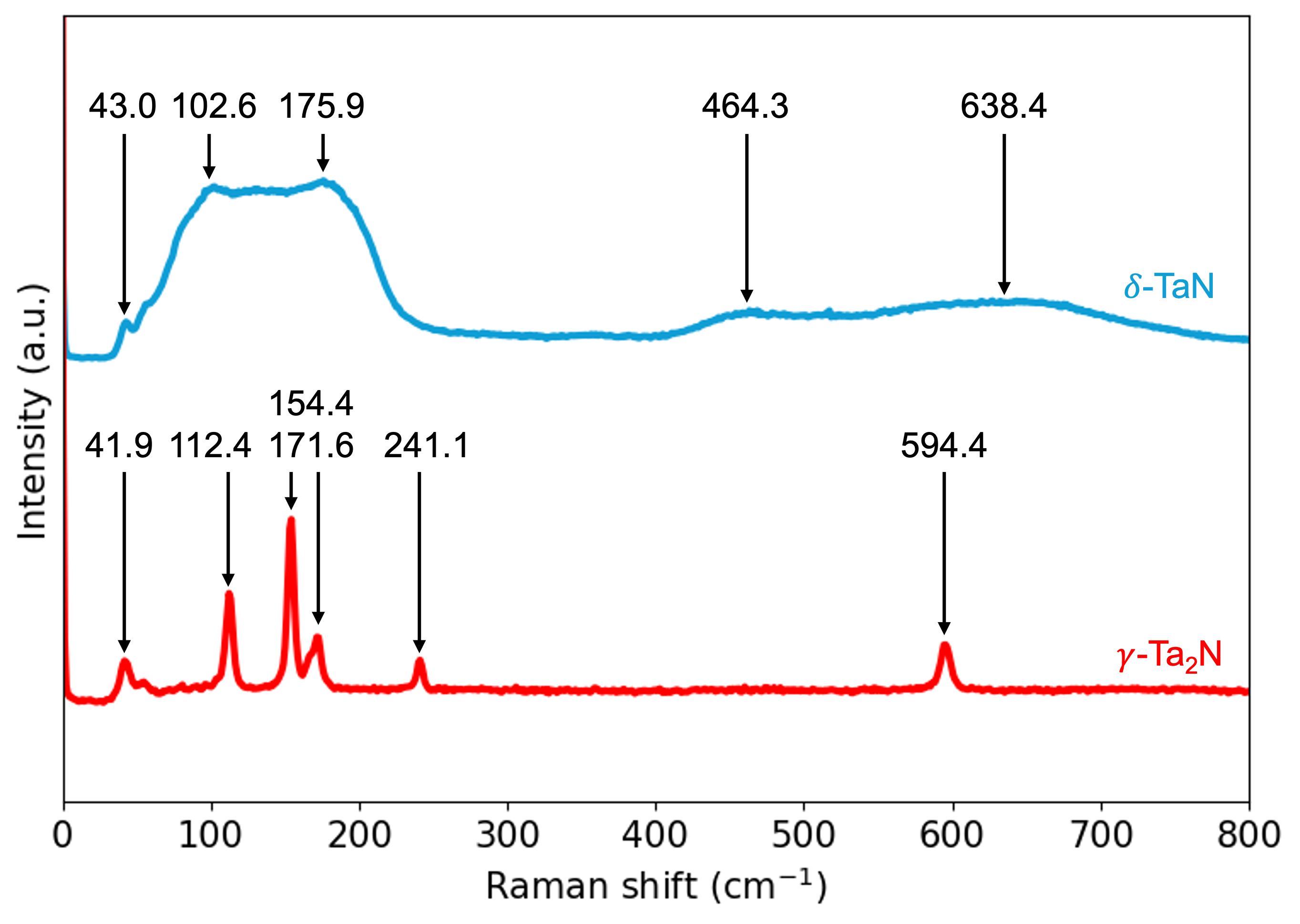}
    \caption{Raman spectra of the 30 nm $\delta$-TaN and 38 nm $\gamma$-Ta$_2$N films. The peaks in Raman spectra, which are unique to each phase of tantalum nitride, can be used to identify the phases present in a film.}
    \label{fig:Raman}
\end{figure}

\begin{figure}
    \includegraphics[width=0.5\textwidth]{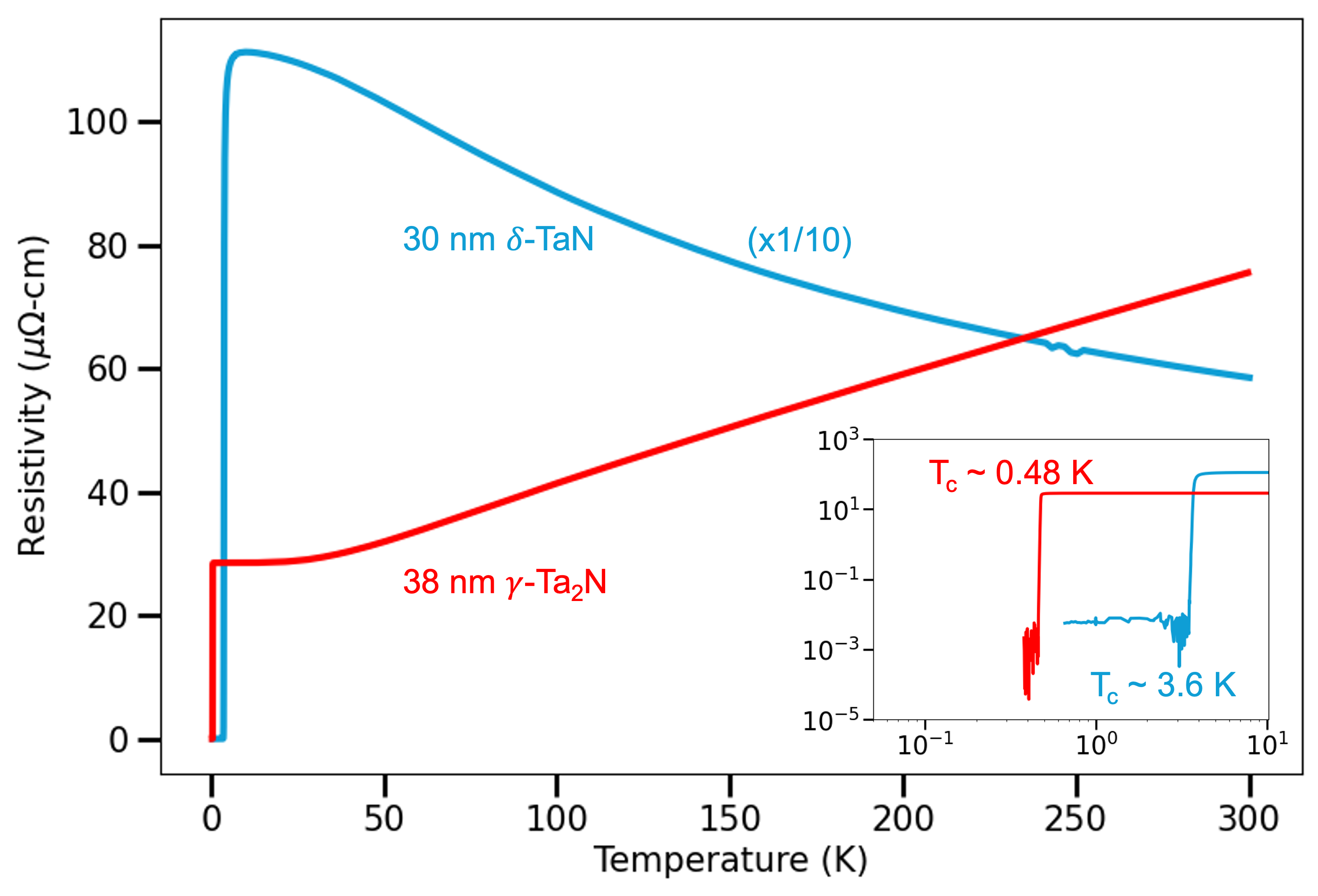}
    \caption{Temperature dependent resistivity ($\rho$ vs T) measurements of the 30 nm $\delta$-TaN and 38 nm $\gamma$-Ta$_2$N films. $\rho$ vs T of 30 nm $\delta$-TaN is multiplied by 1/10 to have the plots in same scale. The actual room temperature resistivities are 586.2 $\mu\Omega$-cm for the 30 nm $\delta$-TaN film and 75.5 $\mu\Omega$-cm for the 38 nm $\gamma$-Ta$_2$N film. (Inset) The log-log plot of $\rho$ vs T showing a superconducting transition at around 3.6 K for the 30 nm $\delta$-TaN film and 0.48 K for the 38 nm $\gamma$-Ta$_2$N film.}
    \label{fig:Tc}
\end{figure}

\begin{figure}
    \includegraphics[width=0.5\textwidth]{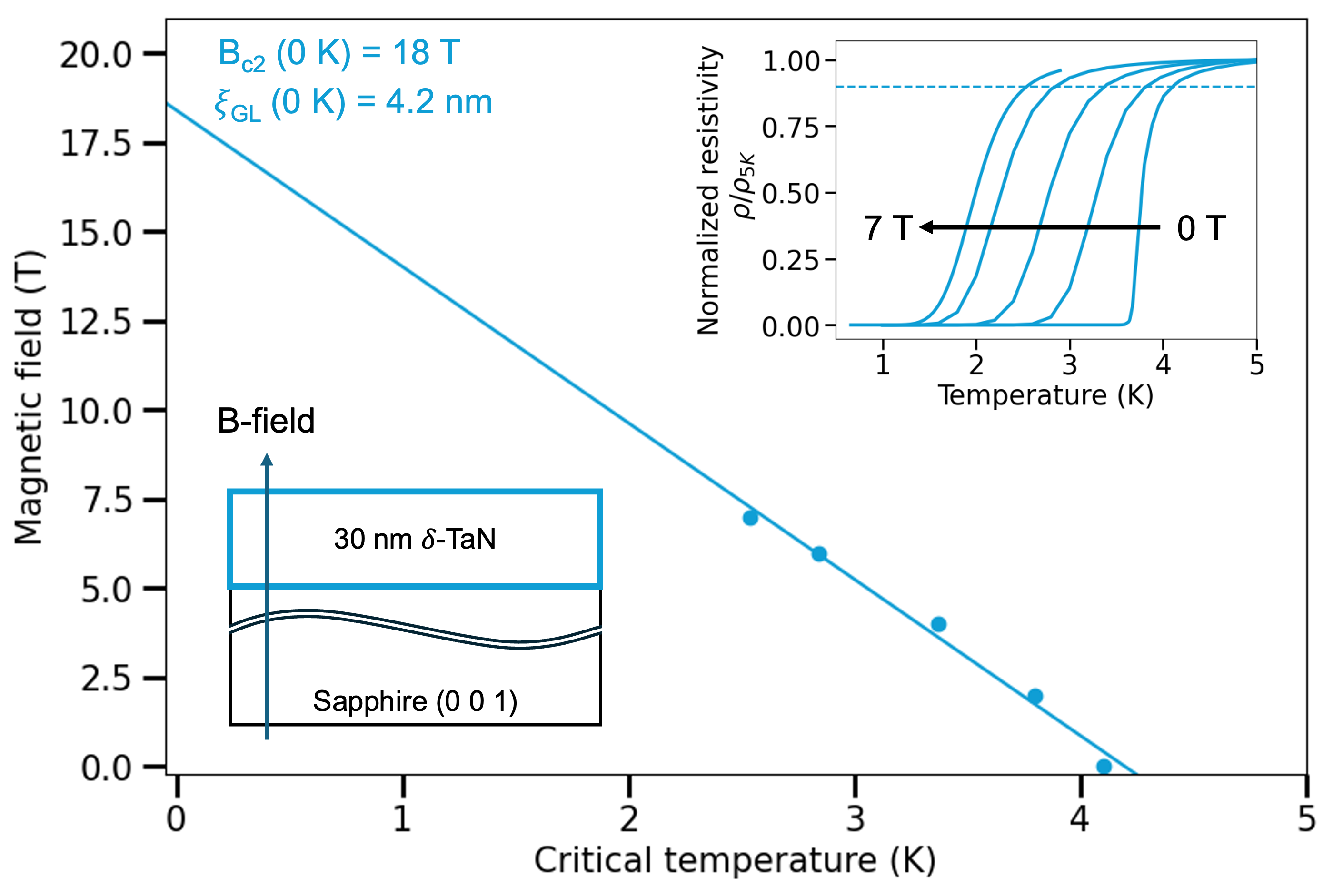}
    \caption{Superconducting transition temperatures of the 30 nm $\delta$-TaN film, taken as the temperatures where the resistance value drops to 90$\%$ of the normal state value at 5 K, extracted at various applied perpendicular magnetic fields [0 T, 2 T, 4 T, 6 T, 7 T] were fit using linearized Ginzburg-Landau (GL) formula. This linear GL fit resulted in 0 K critical field of 18.4 T and GL coherence length of 4.23 nm. (Inset) $\rho$ vs T curves normalized to the $\rho$ value at 5 K.}
    \label{fig:TaN_Bc}
\end{figure}

\begin{figure}
    \includegraphics[width=0.5\textwidth]{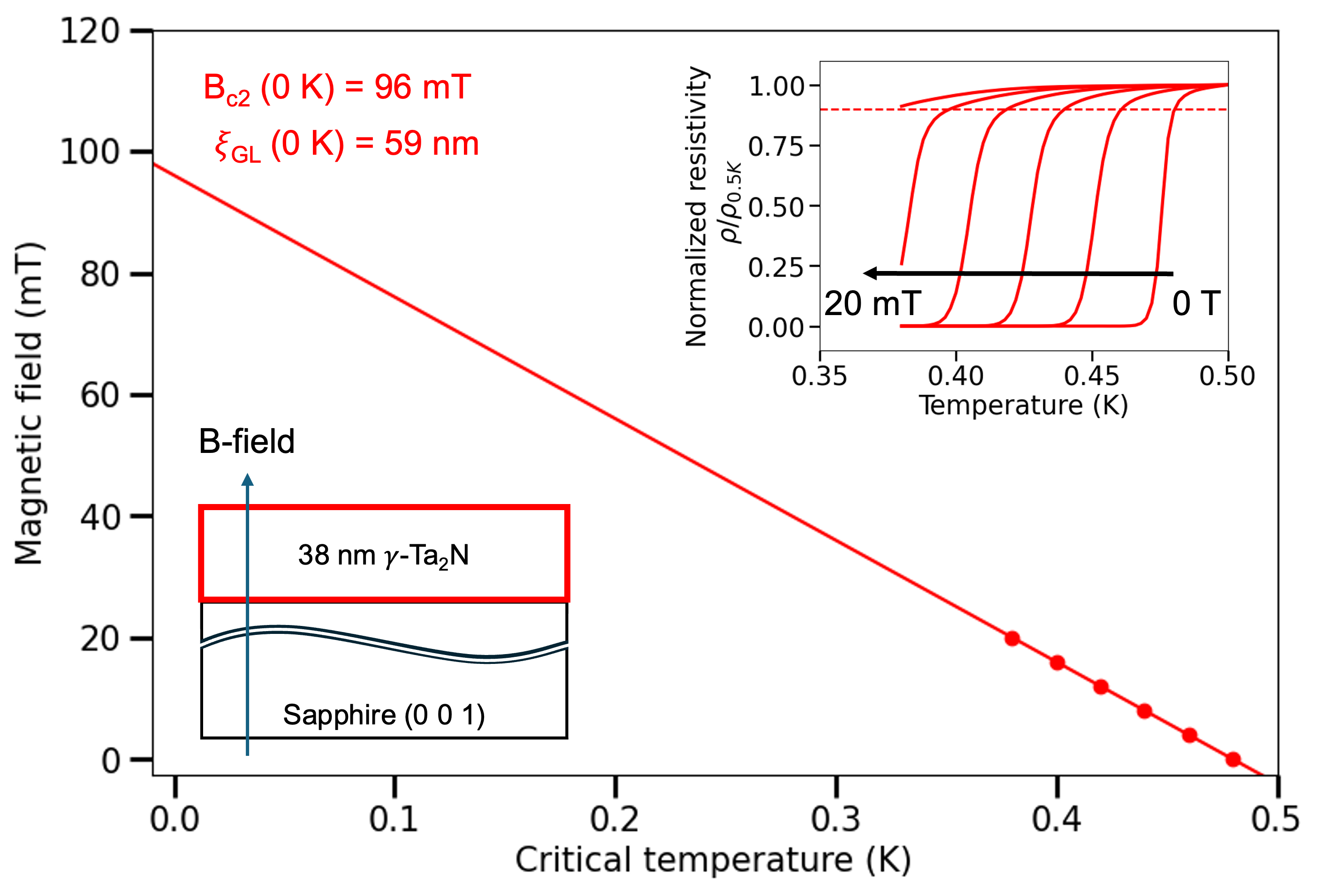}
    \caption{Superconducting transition temperatures of the 38 nm $\gamma$-Ta$_2$N film, taken as the temperatures where the resistance value drops to 90$\%$ of the normal state value at 0.5 K, extracted at various applied perpendicular magnetic fields [0 mT, 4 mT, 8 mT, 12 mT, 16 mT, 20 mT] were fit using linearized Ginzburg-Landau (GL) formula. This linear GL fit resulted in 0 K critical field of 96 mT and GL coherence length of 58.6 nm. (Inset) $\rho$ vs T curves normalized to the $\rho$ value at 0.5 K.}
    \label{fig:Ta2N_Bc}
\end{figure}

Tantalum (Ta) metal has gained significant attention within the superconducting resonator community due to the long lifetimes of about 0.3 milliseconds for Ta-based qubits \cite{0p3msTaQubits} and the high quality factors in Ta co-planar waveguide (CPW) resonators, especially at low photon numbers \cite{DisentanglingTaLosses}. The reason behind the success of Ta in this field is still being investigated. The primary hypothesis is that the native oxide of Ta, Ta$_2$O$_5$, is not only the most stable native oxide of Ta but also aggressive acid treatment can selectively remove this oxide and other contaminants without affecting the Ta metal \cite{TaSurfaceOxideChemistry}. This motivates us to investigate tantalum nitride which should reduce the surface oxide even more due to the presence of Ta-N bonds. To test whether tantalum nitride films can improve the state-of-the-art resonator quality factors, it is important to deposit these films with least amount of defects on a low-dielectric loss substrate like sapphire (Al$_2$O$_3$). Molecular beam epitaxy (MBE) as a deposition technique specializes in these extreme requirements and has also seen success over other deposition techniques in the superconducting resonator field in the recent past \cite{MBEresonatorsMartinisCleland, olson2015growth, MBEresonatorsChrisRichardson, CoreyRaeBenchmark}. With this motivation, we developed the growth conditions for stabilizing single crystal, phase-pure tantalum nitride films on c-plane sapphire by MBE.

The tantalum metal source with a purity of 99.95$\%$ was evaporated using an e-beam source connected to the Veeco GenXplor MBE system with a base pressure below 10$^{-10}$ Torr and the active nitrogen flux was generated using a radio-frequency plasma source connected to a 6N5 pure N$_2$ gas. During growth the chamber pressure was around $2\times10^{-5}$ Torr. The schematics of the two samples used in this study are shown in Fig. \ref{fig:RHEED_AFM}(a) and (b). The sample in Fig. \ref{fig:RHEED_AFM}(a) was grown at a substrate temperature of 600$^\text{o}$C as measured by thermocouple near the heater with nitrogen flow rate of 3 sccm (cubic centimeter per minute at standard temperature and pressure) and a plasma power of 450 W. After a 15 minute growth, the thickness as measured by x-ray reflectivity was 30 nm. Similarly, the sample in Fig. \ref{fig:RHEED_AFM}(b) was grown at a substrate temperature of 1000$^\text{o}$C with a nitrogen flow rate of 2 sccm and plasma power of 200 W. The thickness of this film was determined to be 38 nm.

The reflection high energy electron diffraction (RHEED) patterns of the tantalum nitride films of Fig. \ref{fig:RHEED_AFM}(a) and (b) are shown in Fig. \ref{fig:RHEED_AFM}(c) and (d) respectively. The RHEED patterns indicate that the surface of the sample grown at 600$^\text{o}$C has a twinned cubic structure with 3-dimensional growth mode represented by spotty RHEED pattern. This corresponds to the $\delta$-TaN phase with Fm$\overline{3}$m space group. This is very similar to niobium nitride grown at the same temperature in the same system as reported by Wright \textit{et al}. \cite{NbNonSapphireByJohn}. Similar to the high temperature growths of niobium nitride by Wright \textit{et al}., the RHEED pattern of the film grown at 1000$^\text{o}$C indicates a 2-dimensional growth mode \cite{NbNonSapphireByJohn}. 
This phase was later identified as $\gamma$-Ta$_2$N phase with P$\overline{3}$m1 space group. 
We note that this hexagonal phase is directly analogous to $\beta$-Nb$_2$N, see \cite{NbNonSapphireByJohn,NRLpaperNbNandTaNonSiC} (it is referred to as $\gamma$-Ta$_2$N is because of the presence of a $\beta$-Ta allotrope, the corresponding phase of which is absent for Nb).

Atomic force microscopy (AFM) scans of both the films in Fig. \ref{fig:RHEED_AFM}(a) and (b) shown in Fig. \ref{fig:RHEED_AFM}(e) and (f) respectively indicate that both the films were very smooth with a root-mean-square roughness of less than 0.3 nm. However, the AFM features observed were not in accordance with the corresponding results of niobium nitride from Wright \textit{et al}. \cite{NbNonSapphireByJohn}. For example, the atomic steps were not observed in the high temperature phase of tantalum nitride. The step flow growth mode of $\gamma$-Ta$_2$N phase may be expected at even higher growth temperatures.

Fig. \ref{fig:XRD} shows the symmetric x-ray diffraction (XRD) scans of both $\delta$-TaN and $\gamma$-Ta$_2$N films. The scans were offset for visualization. The XRD data indicates a presence of only one phase in each of these samples. These peak positions were matched with the bulk tantalum nitride XRD data to identify the phase and orientation of the crystals with respect to sapphire substrate. The crystal structures thus identified are represented on the Fig. \ref{fig:XRD} close to the corresponding scan. The (1 1 1) orientation of $\delta$-TaN and (0 0 2) orientation of $\gamma$-Ta$_2$N were identified as expected from the niobium nitride results by Wright \textit{et al}. \cite{NbNonSapphireByJohn}. The Pendell{\"o}sung fringes are clearer and significantly more in the case of $\gamma$-Ta$_2$N compared to that of $\delta$-TaN. This is also consistent with the XRD results from niobium nitride \cite{NbNonSapphireByJohn}. 

Reciprocal space maps (RSMs), as presented in Fig. \ref{fig:RSM}, were used to further determine the phase purity of these films. The scans were all performed with the [1 0 0] axis of sapphire being parallel to the x-ray beam path. This means that the planes of tantalum nitride films and sapphire substrate detected in the measurement share the sapphire [1 $\overline{1}$ 0] zone axis. Fig. \ref{fig:RSM}(a) shows the legend of the planes of each of the phases we looked for. Fig. \ref{fig:RSM}(b) and (c) show the RSMs of the planes shown in Fig. \ref{fig:RSM}(a) for the $\delta$-TaN and $\gamma$-Ta$_2$N films as indicated by the inset schematics.

In Fig. \ref{fig:RSM}(b), only TaN (1 3 3) plane is observed in addition to the sapphire (1 1 9) plane indicating phase-pure $\delta$-TaN. Similarly, only Ta$_2$N (1 0 5) plane is observed in addition to the sapphire (1 1 9) plane in Fig. \ref{fig:RSM}(c) indicating phase-pure $\gamma$-Ta$_2$N. The unit cells and the colored planes shown in the insets of Fig. \ref{fig:RSM}(b) and (c) are based on the crystal symmetry and the measured planes respectively and are not to scale.

Based on the x-ray diffraction data, the lattice constant of $\delta$-TaN film was extracted to be $a = 4.306 \textup{~\AA}$. This is very close to the bulk lattice constant $a = 4.331 \textup{~\AA}$ of rocksalt cubic $\delta$-TaN given by Gatterer \textit{et al}. \cite{TaNlatticeConstant}[ICSD-76456]. Similarly, the lattice constants of $\gamma$-Ta$_2$N are extracted as $a = 3.060 \textup{~\AA}$ and $c = 4.919 \textup{~\AA}$, which also match the bulk $\gamma$-Ta$_2$N lattice constants of $a = 3.0476 \textup{~\AA}$ and $c = 4.9187 \textup{~\AA}$ given by Brauer \textit{et al}. \cite{Ta2NlatticeConstant} [ICSD-76015].

It must be noted that the Ta-N equilibrium phase diagram is very similar to but a little bit more complicated than the Nb-N phase diagram due to increased covalent nature of Ta-N bonds \cite{Terao1971,NRLpaperNbNandTaNonSiC}. This leads to a wide range of Ta:N stoichiometries being stable with different crystal structures. Each of these crystal structures will have their own orientation corresponding to the sapphire substrate. In Fig. \ref{fig:RSM}, we only looked for the tantalum nitride phases whose orientations with respect to sapphire were understood from niobium nitride phases. Nevertheless, the singular peaks in XRD scans shown in Fig. \ref{fig:XRD} along with RSMs shown in Fig. \ref{fig:RSM} indicate phase purity in both the films.

Fig. \ref{fig:Raman} shows the Raman spectra of the $\delta$-TaN and $\gamma$-Ta$_2$N films collected using a WITec alpha300 Raman microscope with a pump laser wavelength of 532 nm at room temperature. Raman spectra of superconducting refractory metals has long been investigated to understand the relationship between phonon density and superconducting transition temperature (see for example \cite{NbNxRaman}). Such studies in niobium nitride revealed a distinct Raman signature associated with each phase \cite{NbNxRaman}. Raman spectroscopy can thus be used to determine phase purity of samples from material systems with complex phase diagrams such as tantalum nitride and niobium nitride. As can be seen in Fig. \ref{fig:Raman}, our $\delta$-TaN and $\gamma$-Ta$_2$N films also exhibit significantly different Raman spectra, given the differences in both crystal structure and stoichiometry between the phases. The Raman spectra and the corresponding peak positions that we report here (Fig. \ref{fig:Raman}) may be used as reference for phase pure $\delta$-TaN and $\gamma$-Ta$_2$N. Note that the broad band of peaks between 100 and 200 cm$^{-1}$ for $\delta$-TaN is similar to that of $\delta$-NbN reported in \cite{NbNxRaman}.

The temperature-dependent electrical resistivity of $\delta$-TaN and $\gamma$-Ta$_2$N films was measured using a four probe method in a Quantum Design Dynacool physical properties measurement system (PPMS). An additional He-3 option, capable of reaching 0.36 K, was used to observe superconductivity in $\gamma$-Ta$_2$N. The resistivity vs temperature data is shown in Fig. \ref{fig:Tc}. The resistivity of the $\delta$-TaN film is 10 times the value indicated by the y-axis. This scaling allows us to compare both the films in one plot. 
A direct comparison to the literature for the $\delta$-TaN film shown in this work is difficult due to its novelty and differences in the thickness, deposition technique, substrate etc. However, the reported room temperature resistivity values in the literature are about 1000 $\mu\Omega$-cm \cite{VanDuzerTaNresistivity,TaNresistivityAboutAmilliOhm}. This is in agreement with the room temperature resistivity of our 30 nm $\delta$-TaN film which is 586.2 $\mu\Omega$-cm. The resistivity of our $\delta$-TaN film is slightly lower possibly due to improved crystalline quality and chemical purity. On the other hand, the room temperature resistivity of our 38 nm $\gamma$-Ta$_2$N film is 75.5 $\mu\Omega$-cm and is comparable to the existing literature \cite{NRLpaperNbNandTaNonSiC}.

The residual resistivity ratio (RRR), defined as $\rho_{300\text{ K}}/\rho_{10 \text{ K}}$, of the $\delta$-TaN film is 0.52, which is less than unity, similar to the RRR of $\delta$-NbN as reported by Wright \textit{et al}. \cite{NbNonSapphireByJohn}. For the $\gamma$-Ta$_2$N film, the RRR is 2.6, similar to the RRR of $\beta$-Nb$_2$N \cite{NbNonSapphireByJohn}. The inset of Fig. \ref{fig:Tc} shows the superconducting transition in both the films. $\delta$-TaN has a superconducting transition temperature ($T_\text{c}$) $\simeq$ 3.6 K,  although $\delta$-TaN has been shown to have $T_\text{c}$ as high as 10.8 K in a 300 nm TaN film sputter deposited on molybdenum substrate at room temperature \cite{TaNcriticalTemperature10p8K}. This is possibly due to low temperature growth, lower thickness and phase purity of our film. The mixed phases have been shown to possess the highest $T_\text{c}$ in the niobium nitride family on both sapphire and SiC substrates \cite{NbNonSapphireByJohn,NbNonSiCbyJohn}. Moreover, the $T_\text{c}\simeq 3.6$ K of our 30 nm $\delta$-TaN film is very close to the $T_\text{c}\simeq 3.7$ K of $\delta$-TaN film of similar thickness reported previously \cite{36nmTaNwithTc3p7Krho1800muOhmcm}. $\gamma$-Ta$_2$N has a $T_\text{c}$  $\simeq$ 0.48 K very similar to that of $\beta$-Nb$_2$N \cite{NbNonSapphireByJohn}. This is the first time a phase pure $\gamma$-Ta$_2$N has been shown to be superconducting. The previous MBE-grown $\gamma$-Ta$_2$N by Katzer \textit{et al}., was shown to have higher $T_\text{c}$ at an expense of phase mixture \cite{NRLpaperNbNandTaNonSiC}.

We measured the $T_\text{c}$ under a range of magnetic fields perpendicular to the surface of both the films as shown in Figs. \ref{fig:TaN_Bc} and \ref{fig:Ta2N_Bc}. The $T_\text{c}$ was extracted to be the temperature at which the resistance drops to 90\% of the normal state resistance just above the transition temperature. The magnetic-field-dependent normalized $\rho$ vs T curves and the corresponding 90\% lines are shown as insets in Figs. \ref{fig:TaN_Bc} and \ref{fig:Ta2N_Bc}. The linearized Ginzburg-Landau (GL) fit is given by 

\begin{equation}
    B_{\text{c}2}^{\perp}(T) = \frac{\phi_0}{2\pi\xi^2}\left(1-\frac{T}{T_\text{c}}\right),
    \label{Eq. LinearGL}
\end{equation}
where $\phi_0 = h/2e$ is the magnetic flux quantum, a ratio of the fundamental constants - Planck's constant $h$ and fundamental charge $e$. The factor of 2 in front of $e$ is due to the Cooper pairs in superconductors. $B_{\text{c}2}^\perp(T)$ is the upper critical field perpendicular to the film at a given temperature $T$, and $\xi=\xi_{\text{GL}}(0 \text{ K})$ is the GL coherence length at 0 K.

Eq. \ref{Eq. LinearGL} was then used to extract the 0 K GL coherence length, $\xi_{\text{GL}}$ and the 0 K critical magnetic field $B_{\text{c}2}$ similar to Yan \textit{et al}. \cite{RusenYanNature}. Both the linear GL fits shown in Figs. \ref{fig:TaN_Bc} and \ref{fig:Ta2N_Bc} had $>$ 99\% r-squared value indicating a good fit. The extracted values are $\xi_{\text{GL}}(0 \text{ K})\simeq 4.2$ nm and $B_{\text{c}2}(0 \text{ K})\simeq 18$ T for 30 nm $\delta$-TaN film. The previously reported $B_{\text{c}2}(0 \text{ K})$ of 13.8 T matches closely with our results \cite{TaNcriticalfield14T}. Similar to the structural properties, the $B_{\text{c}2}(0 \text{ K})$ of 30 nm TaN film matches that of the values of delta niobium nitride thin film with much higher $T_\text{c}\simeq$ 16 K reported by Dang \textit{et al}. \cite{PhillipDangNbNcriticalfield}. $\xi_{\text{GL}}(0 \text{ K})\simeq 59$ nm and $B_{\text{c}2} (0 \text{ K})\simeq 96$ mT extracted for 38 nm $\gamma$-Ta$_2$N film is the first measurement of this kind on this phase to the best of our knowledge and hence cannot be compared with literature values. GL analysis was also performed with critical temperature defined by 10\% of the normal state resistance just above the transition temperature. This resulted in a 13 T critical magnetic field, 5.1 nm GL coherence length for the $\delta$-TaN film and 77 mT critical magnetic field, 66 nm GL coherence length for the $\gamma$-Ta$_2$N film.  

In conclusion, successful growth conditions were identified for obtaining two single crystal, phase-pure tantalum nitride phases on c-plane sapphire by tuning the growth temperature and the Ta metal flux. The phase purity of these films was identified from x-ray measurements. The temperature-dependent electronic measurements were then performed on these films to identify crucial superconducting parameters like critical temperature, critical magnetic field, and GL coherence lengths. 
Moving forward, we anticipate realization of superconducting resonators from MBE-grown phase-pure single crystal tantalum nitride films. Theoretical and experimental assessment of the evolution of Ta-N bonds at the surfaces in the presence of atmospheric oxygen will greatly support future work on the tantalum nitride platform.

This work is supported by the AFOSR/LPS program Materials for Quantum Computation (MQC) as part of the EpiQ team monitored by Dr. Ali Sayir of AFOSR and Dr. Erin Cleveland of LPS, and partially by an ONR Grant \# N00014-22-1-2633 monitored by Dr. Paul Maki. This material is based upon work supported by the Air Force Office of Scientific Research under award number FA9550-23-1-0688. Any opinions, findings, and conclusions or recommendations expressed in this material are those of the author(s) and do not necessarily reflect the views of the United States Air Force. This work was performed in part at the Cornell NanoScale Facility, a member of the National Nanotechnology Coordinated Infrastructure (NNCI), which is supported by the National Science Foundation (Grant NNCI-2025233). The authors acknowledge the use of facilities and instrumentation supported by NSF through the Cornell University Materials Research Science and Engineering Center DMR-1719875.

\clearpage




\bibliography{SingleXtalTaN}

\end{document}